\documentclass[aps,pra,a4paper,twocolumn,longbibliography,superscriptaddress]{revtex4-2}
\usepackage[utf8]{inputenc}
\usepackage{amsmath,amsfonts,amssymb}
\usepackage{mathtools}
\usepackage{graphicx,xcolor}
\usepackage[space]{grffile}   
\usepackage[normalem]{ulem}     

\newcommand{\beq}{\begin{equation}}
\newcommand{\eeq}{\end{equation}}
\newcommand{\bei}{\begin{itemize}}
\newcommand{\eei}{\end{itemize}}
\newcommand{\ben}{\begin{enumerate}}
\newcommand{\een}{\end{enumerate}}

\newcommand{\be}{{\mathbf e}}

\usepackage{hyperref}
\hypersetup{
   pdfpagemode=UseNone, 
   pdfstartpage=1,
   pdfmenubar=true,
   pdftoolbar=true,
   colorlinks=true,
   linkcolor=blue,
   citecolor=blue,
   urlcolor=blue,
   bookmarksopen=false
 }

\definecolor{darkblue}{rgb}{0.,0.24,0.51}
\definecolor{britishracinggreen}{rgb}{0.0, 0.26, 0.15}
\definecolor{darkgreen}{rgb}{0,0.60,.2}

\def\be{\begin{equation}}
\def\ee{\end{equation}}

\def\rf#1{(\ref{#1})}

\begin{document}
\title{
Exact results for heavy unitary Bose polarons
}
\author{Nikolay Yegovtsev}
\affiliation{Department of Physics and Center for Theory of Quantum Matter, University of Colorado, Boulder CO 80309, USA}
\author{Grigori E. Astrakharchik}
\affiliation{Departament de F\'isica, Universitat Polit\`ecnica de Catalunya, Campus Nord B4-B5, E-08034 Barcelona, Spain}
\author{Pietro Massignan}
\affiliation{Departament de F\'isica, Universitat Polit\`ecnica de Catalunya, Campus Nord B4-B5, E-08034 Barcelona, Spain}
\author{Victor Gurarie}
\affiliation{Department of Physics and Center for Theory of Quantum Matter, University of Colorado, Boulder CO 80309, USA}

\date{\today}
	
\begin{abstract}
We consider the problem of unitary Bose polarons, i.e., impurities interacting  via a potential with infinite scattering length with a bath of  weakly interacting bosons.  
We provide an analytic expression for the energy of a heavy impurity whose interaction potential has a range larger than the healing length of the bath.
Furthermore, we perform numerically exact Diffusion Monte Carlo calculations and we demonstrate that the simple Gross-Pitaevskii theory provides a remarkably accurate description of heavy unitary Bose polarons throughout the whole experimentally relevant range of gas densities.
\end{abstract}
\maketitle

\section{Introduction}
The study of dilute quantum impurities immersed in a bath of identical bosons or fermions has been a topic of significant interest over the past few years \cite{Chevy2010,Massignan_Zaccanti_Bruun,Schmidt2018,Scazza2022,Parish2023}. 
The main driver of such efforts was the tremendous experimental progress in ultracold atomic gases and atomically-thin semiconductors~\cite{Schirotzek2009,Kohstall2012,Koschorreck2012,Jorgensen2016,Hu2016,Scazza2017,Sidler2017,Yan2019,PenaArdila2019,Yan2020,Tan2020,Skou2021,Fritsche2021,Muir2022,Tan2022,Huang2023,Baroni2023,Morgen2023}, where researchers learned how to control with great precision the strength of the interaction between the dilute impurities and the bath, the temperature of the system, and even change the mass or the quantum statistics of the impurities.

In particular, it is nowadays feasible to investigate theoretically and experimentally the scenario of strong (``unitary") interactions in which the scattering length of the impurity-bath two-body potential is tuned to infinity, and therefore the physics becomes independent of it.
If the number of remaining relevant scales in the problem is small, one anticipates that the properties of the system can be described by universal combinations of those. What makes the unitary polaron especially intriguing is that while the results are independent of the bath-impurity scattering length, the problem remains nontrivial because of the strong interactions. In this scenario, the physical intuition built on perturbation theory dramatically fails, and one has to rely on numerical methods.

An example of particular interest is that of an impurity immersed into a weakly interacting BEC, which is commonly referred to as the Bose polaron problem \cite{Gross1962,Padmore1971,Astrakharchik2004,Massignan2005,Rath2013,Shashi2014,Casteels2014,Christensen2015,Ardila2015,Shchadilova2016,Drescher2020,Levinsen2021,Schmidt2021,seetharam2021quantum}. Most of the recent theoretical studies assumed that adding a single impurity will not significantly distort the bath density, and therefore expanded the problem around the translationally-invariant ground state of the interacting bosons in the absence of the impurity. However, even a single impurity can significantly affect the density of a dilute bosonic bath (due to the large compressibility of the latter), and indeed a true ``orthogonality catastrophe" arises when the bath becomes an ideal Bose gas \cite{Massignan2005,Yoshida2018,Guenther2020}. 
As such, the prediction of the energy of the polaron computed starting from an unperturbed bath turns out to be incorrect, as has been recently shown by using more advanced analytical methods and performing Monte Carlo simulations \cite{Levinsen2021,Massignan2021,Yegovtsev2022}. 
A natural way to circumvent this issue is therefore to solve the problem in real space by means of the Gross-Pitaevskii equation~(GPe), which self-consistently deforms the density profile of the bath around the impurity, and in particular captures correctly the emergence of the orthogonality catastrophe \cite{Guenther2020}. This approach has been particularly fruitful in the study of 1D Bose polarons with contact interactions, where the problem becomes integrable and one can even go beyond mean-field and account for quadratic fluctuations, showing a good agreement with the Quantum Monte Carlo studies \cite{jager2020strong,petcovich2103mediated, will2021polaron}. Similarly, in 3D, if the gas parameter remains small everywhere, even in the vicinity of the impurity where the bath density may be highest, the mean-field GP picture will provide an accurate description of the Bose-polaron properties. 
Such mean-field studies have been performed before, but the main emphasis was on the numerics \cite{Massignan2005, Guenther2020, Drescher2020, hryhorchak2020impurity,Schmidt2021}.
Quite remarkably, the GPe for a heavy impurity with a finite range of the boson-impurity potential, infinitesimally short range of the boson-boson potential, and sufficiently low gas density may be solved analytically even for strong interactions (at unitarity, and in its neighborhood), as we have shown in Refs.~\cite{Massignan2021,Yegovtsev2022}.

While it is known that the GPe correctly reproduces the properties of the heavy polarons in the regime of weak boson-impurity interactions \cite{Astrakharchik2004, Guenther2020, Yegovtsev2022}, a recent Monte Carlo study by \cite{Levinsen2021} argued that this approach should break down at unitarity. Their critique of the GP approach is based on the statement that the number of particles trapped within an impurity should be much larger than unity, and together with the smallness of the gas parameter, this should imply $a_b\ll r_c$, where $a_b$ is the boson-boson scattering length and $r_c$ is the characteristic size of the impurity potential. The same condition was first derived in \cite{chen2018trapping} for the applicability of the GPe in the presence of impurity-bath bound states. However, here we formally consider the unitary case where the impurity-bath bound state is absent, so we believe the above condition does not have to be satisfied. This is confirmed by the Monte Carlo numerics that will be presented throughout this paper. 
Another claim of \cite{Levinsen2021} is that the energy of the polaron as a function of the gas parameter has a universal form containing a logarithm.  This prediction differs from what was found in \cite{Massignan2021,Yegovtsev2022}. However the model potentials used by \cite{Levinsen2021} are not compatible with a GP treatment of the problem, so a direct comparison is not possible. 
This prompted us to perform a separate Monte Carlo study where we chose the boson-boson and boson-impurity potentials in a way that allows us to study the problem using the GPe and the Monte Carlo methods on equal footing, for which such a comparison can be made.

To test whether the solution of the GPe correctly captures the behavior of the unitary Bose polaron in 3D, we perform here an extensive numerical study of the Bose polaron at unitarity using both the GPe and the Diffusion Monte Carlo (DMC) methods. In particular, we study the effects of including a non-zero range of the boson-boson interaction potential and the applicability of the Born approximation to this potential, and we demonstrate the remarkable accuracy of the simple GP approach. In addition, we find an analytic solution of the problem based on the Local Density Approximation which is valid when the impurity-bath potential has a range larger than the healing length of the bath, which is typically the case for ionic and Rydberg impurities \cite{durst2024phenomenology}.

This paper is organized as follows. In Section~\ref{sec:setup} we introduce the Hamiltonian of the system of interest together with the model potentials that characterize the boson-boson and boson-impurity interactions that are used in our numerical simulations. Here we also discuss the role of different length scales in the problem and show that at unitary the local GPe is governed by a single dimensionless parameter $\epsilon = r_c/\xi$, the ratio of the characteristic size of the boson-impurity potential and healing length of the BEC. In Section~\ref{sec:analytics} we show that this local GPe can be analytically solved in two distinct regimes. Solution in the regime $\epsilon\ll1$ has been previously discussed in Refs.~\cite{Massignan2021,Yegovtsev2022}, so we briefly review its properties for later convenience. Additionally, we provide a simple derivation of the solution in the regime $\epsilon\gg1$ together with the corresponding expressions for the energy of the polaron and the number of trapped particles in the polaronic cloud. We also compare our results with the phenomenological expression for the energy of the polaron proposed in Ref.~\cite{Schmidt2021} and show that contrary to their prediction, the energy of the polaron does not depend on the effective range of the boson-impurity potential. Finally, we argue that the effects of finite boson-boson range are typically small and can be effectively incorporated into the above picture using the first order perturbation theory within the generalized GPe. For the Bose polaron at unitarity this produces a nonuniversal small negative shift in energy. In Section~\ref{sec:numerics} we present our numerical results based on local GPe, non-local GPe and the DMC and show the consistency of the results across all three methods. We also show that our analytical results are validated by the numerical analysis. Section~\ref{sec:conslusions} presents our conclusions. Finally, in the Appendices we discuss the applicability of the Born approximation to the boson-boson potential and solve a toy model of a square well potential at weak coupling that shows that unless $\epsilon\ll1$, the solution depends on various details of the potential, so the properties of the polaron in this regime are not universal. The details of the numerical simulations are also presented there.

\section{Problem set up}
\label{sec:setup}
We consider bosons of mass $m$ at density $n_0$ interacting among themselves via a short-range potential $V_{bb}$. Additionally, bosons interact with a single infinitely massive (i.e. pinned) impurity via the potential $U$, which effectively acts as an external field. 
The microscopic Hamiltonian describing the Bose polaron problem reads
\be \label{eq:ham}
H =-\sum\limits_i \frac{\Delta_i}{2m} 
+ \sum\limits_{i<j} V_{bb}({\bf x}_i-{\bf x}_j) + \sum\limits_i U({\bf x}_i-{\bf X}),
\ee
where ${\bf x}_i$ represents the coordinates of the bosons, ${\bf X}$ denotes the position of the impurity, and we have set $\hbar=1$. 
To find the ground-state energy of this Hamiltonian we follow various strategies. 
On the one hand, we perform DMC simulations for a fixed number of particles (typically, 100).
Next to that, we solve the problem at the mean-field level by means of the GPe
\be
\label{eq:GPnl} \left[ - \frac{\Delta}{2m} + U({\bf x})  -  \mu  + \int d^3 {\bf y} \ V_{bb}({\bf x-y}) \left| \psi({\bf y}) \right|^2 \right] \psi({\bf x}) = 0,
\ee
where the particle density is now fixed by the chemical potential $\mu$, and without loss of generality we located the impurity at the origin, ${\bf X}=0$.

Solving the {\it non-local GPe}~(\ref{eq:GPnl}) poses serious analytical and numerical challenges due to its non-linear integro-differential nature. Choosing a potential between bath particles like
\be 
\label{eq:vbb1} 
V_{bb}({\bf x-y}) = \frac{\lambda}{\pi^{3/2}r_b^3}\,e^{-|{\bf x-y}|^2/r_b^2}
\ee
permits to reduce Eq.~\rf{eq:GPnl} to a one-dimensional equation in the radial direction, simplifying drastically the complexity of its numerical solution~\cite{Yegovtsev2022}. The parameters of the above potential are chosen in a such way that in the zero-range limit, $r_b \to 0$, one formally recovers the usual {\it local GPe}, introduced in Eq.~\eqref{eq:GPl} below. 
The $s$-wave scattering length $a_b$ of this potential is determined by the solution of the zero-energy scattering problem and depends on both the amplitude $\lambda$ and the range $r_b$. 
In the first Born approximation, applicable for $a_b \ll r_b$, the scattering length is directly proportional to the amplitude of the Gaussian potential~(\ref{eq:vbb1}), $a_b = \lambda m/(4 \pi)$.
The situation in which the potential range $r_b$ is of the order of the scattering length $a_b$
is a delicate one and will be discussed in detail below [see also Appendix \ref{sec:born}].

Let us now consider the boson-impurity potential $U(r)$. Various earlier numerical studies focused on short-range interactions represented by a contact pseudopotential. While this choice can be tackled by the DMC method, it is still a very singular limit for the non-local GPe~\rf{eq:GPnl}, which suffers from a $1/r$ divergence at the origin~\cite{Drescher2020}. 
As a result, in the vicinity of the impurity the gas parameter $|\psi({\bf x} \sim 0)|^2a_b^3$ grows indefinitely, leading to a breakdown of the GPe description.
Such a divergence can be avoided by making the range of the impurity potential finite. Then, the predictions of the GPe can be trusted provided that the gas parameter remains small everywhere, including in the vicinity of the impurity~\cite{Massignan2021,Yegovtsev2022}. 
In the following, we model the boson-impurity potential by a Pöschl-Teller potential of range $r_c$ and amplitude tuned to unitarity:
\be
\label{eq:PT}
U(r) = -\frac{1}{mr_c^2 \cosh^2(r/r_c)}.
\ee
At zero energy, the Schr\"odinger equation with this potential has the simple analytic solution $\psi(r)=\tanh(r/r_c)$.

A deeper insight into the underlying physics can be gained by examining the length scales inherent to the model: $n_0^{-1/3}$, $a_b$, $r_c$ and $r_b$, where $n_0$ is the uniform density of the condensate in the absence of the impurity. An additional simplification arises when the boson-boson interactions are replaced by a zero-range potential,
\be \label{eq:vbb0}
V_{bb}({\bf x-y}) = \lambda \, \delta({\bf x-y}),
\ee 
with $\lambda=4\pi a_b/m$.
The choice of a contact interaction potential yields the usual {\it local GPe}
\be
\label{eq:GPl} \left(- \frac{\Delta }{2m} + U  -  \mu  + \lambda | \psi({\bf x}) |^2 \right)  \psi({\bf x}) = 0.
\ee
The local version of the GPe is easier to handle numerically and it also turns out to be analytically tractable in several regimes that we discuss below. 
We show below that replacing $V_{bb}$ by Eq.~\rf{eq:vbb0} only mildly
affects the properties of the Bose polarons that we study. The only relevant length scales left in the microscopic Hamiltonian are $n_0$, $a_b$, and $r_c$. Also, it can be anticipated that the healing length $\xi$ becomes an important parameter of the system. Physically, such a length corresponds to the energy scale fixed by the chemical potential, which in a weakly interacting Bose gas is given by $\mu = 1/(2m\xi^2)$. Scaling the condensate function by its long-range asymptotic value ($\phi = \psi/\sqrt{n_0}$), using $r_c$ as the unit of distance ($y=r/r_c$), and introducing the dimensionless parameter $\epsilon = r_c/\xi$, the local GPe can be brought to the compact form:
\be \label{eq:GPe0}
-\nabla^2\phi + 2mr_c^2U(y)\phi  =\epsilon^2\left(\phi-\phi^3 \right)
\ee
For the unitary boson-impurity potential introduced in Eq.~\eqref{eq:PT}, the left-hand-side of Eq.~\rf{eq:GPe0} becomes independent of $mr_c^2$. Consequently, the only dimensionless parameter that governs the local GP problem is $\epsilon$. 

\section{Analytically tractable regimes of the local GPe at unitarity}
\label{sec:analytics}
Remarkably, Eq.~\rf{eq:GPe0} can be analytically solved in two distinct regimes. Previously we have found the solution in the regime of low gas densities, which for a fixed $a_b$ corresponds to the condition $\epsilon \ll1$. Here we also report the solution to the problem in the opposite regime, $\epsilon \gg1$. In this Section we first review the main features of the analytical solution at low gas density we derived in Refs.~\cite{Massignan2021,Yegovtsev2022}, then we present the solution at high gas density, and finally we consider the importance of including in the treatment the range of the potential between bath atoms

\subsection{Analytical solution of the unitary polaron at low gas densities}
In the case where $\epsilon \ll 1$ the solution at the location of the impurity scales as $\phi(0) \sim 1/\epsilon^{2/3}$. For simplicity, we first consider impurity-bath potentials $U$ that vanish identically beyond some range $r_c$.

The solution of the local GPe~\rf{eq:GPe0} for the case of a weak potential ($|a|^3 \ll \xi^2 r_c$) reads: 
\be \label{eq:fullsolw}  \phi(r)  \approx \left\{  \begin{matrix} \frac{r_c}{r}  \left( 1 - \frac a {r_c} \right)  v\left(\frac r {r_c} \right), & r < r_c , \cr 1 - \frac a r  e^{- \frac{\sqrt{2} r}{\xi}},  & r>r_c .   \end{matrix}
\right. \ee
Here $v$ is the solution to the zero energy Schr\"odinger equation in the potential $U(y)$ with boundary conditions $v(0)=0$ and $v(1)=1$, and $a$ is the corresponding scattering length.

When the potential $U(y)$ is tuned to unitarity, the result becomes:
\be \label{eq:fullsols}  \phi(r)  \approx \left\{  \begin{matrix}  \frac{\xi^{2/3} R^{1/3} }{r}  \, v\left(\frac r {r_c} \right), & r < r_c , \cr 1+\frac{\xi^{2/3} R^{1/3} }{r}   e^{- \frac{\sqrt{2} r}{\xi}},  & r>r_c .   \end{matrix}
\right. \ee
The length $R$ is defined as $R^{-1} = \int_0^\infty dy\, \frac{v(y)^4}{y^2}$, and generally $R\sim r_c$. The energy of the polaron at unitarity reads:
\be
\label{eq:Eunit}
E = - \frac{ \pi n_0 \xi}{m} \left( 3 \delta^{\frac 1 3} - 2 \sqrt{2} \delta^{\frac 2 3} + 4 \delta \ln \delta + \dots \right),
\ee
where $\delta = R/\xi$. Correspondingly, the number of bath particles in the cloud of a unitary polaron is
\be
\label{eq:Nunit}
N = 4 \pi n_0 \xi^3 \left(  \delta^{1/3} - \frac{5}{3 \sqrt{2}} \delta^{2/3} + 2 \delta \ln \delta  + \dots \right).
\ee
The above results for the energy and the number of particles are also applicable to the short ranged potential that do not vanish identically beyond $r_c$. In order to compute $R$ for such potentials, one has to use the boundary conditions $v(0)=0$ and $v(\infty)=1$ for the corresponding zero energy  Schr\"odinger equation. For the unitary Pöschl-Teller potential we have $v(y) = \tanh(y)$, and $R$ is almost equal to the potential range, $R \approx 1.049r_c$. This gives $\delta \approx 1.049 \, \epsilon$, and so the above expansion works in the regime where $\delta \ll 1$.

In the limit of small bath densities we can retain only the leading terms in these expressions and obtain:
\be
\label{eq:Eunit1}
E = - \frac{ 3\pi n_0 \xi \delta^{\frac 1 3} }{m}  = - \frac{3 (\pi n_0)^{2/3}}{2 m} \left( \frac{R}{a_b} \right)^{1/3},
\ee
\be
\label{eq:Nunit1}
N = 4 \pi n_0 \xi^3 \delta^{1/3} =\frac{R^{1/3}}{4 (\pi n_0 a_b^4)^{1/3}}.
\ee
The above results hold provided that both $\delta$ and the value of the local gas parameter at the position of the impurity $n_0a_b^3/\delta^{4/3}$ are much smaller than one. This can be conveniently expressed as:
\be
\left(n_0a_b^3 \right)^{1/4}\ll\frac{R}{a_b}\ll\frac{1}{\sqrt{n_0a_b^3}}.
\ee
When $R$ violates the rightmost condition, one has to solve the GPe either numerically or using the local density approximation as discussed in the next section.



\subsection{Local density approximation}
An analytic solution for Eq.~\rf{eq:GPe0} may also be found  in the opposite limit $\epsilon \gg 1$ (while $n_0 a_b^3$ still remains small), making use of the local density approximation (LDA). The leading terms in this equation are those on its right-hand side (i.e., the ones multiplied by $\epsilon^2$). This tells us that as a first approximation $\phi \approx 1$. 
We seek a solution of the form $\phi = 1 + u_1/\epsilon^2 + u_2/\epsilon^4\cdots$. For convenience, we write $2mr_c^2U(y) = \mathcal{U}(y)$.  
At order $\epsilon^0$ we get: $\mathcal{U}(y) = -2u_1$, from which we obtain $u_1 = -\mathcal{U}(y)/2$. 
This produces $\phi = 1-\mathcal{U}(y)/(2\epsilon^2)$.
The next equation is at the $1/\epsilon^2$ order and it defines $u_2$. However, as will be shown below, the term $u_2$ contributes at order higher than $1/\epsilon^4$, so we do not need it here. 
Indeed, if we plug the solution $\phi = 1 + u_1/\epsilon^2 + u_2/\epsilon^4 \cdots$ into the energy functional we obtain
$$
E = \frac{n_0r_c}{2m}\int d^3{\bf y}\left((\nabla\phi)^2 + (\mathcal{U}(y)-\epsilon^2)\phi^2 + \frac{\epsilon^2}{2}\phi^4 + \frac{\epsilon^2}{2} \right),
$$
Then the term at order $1/\epsilon^4$ that involves $u_2$ vanishes identically, leaving only $(\nabla u_1)^2$ at $1/\epsilon^4$ order. Plugging $u_1 = -\frac{\mathcal{U}(y)}{2}$ into the remaining terms produces the final result:
\begin{equation}
\label{eq:ELDA}
E = \frac{n_0r_c}{2m}\int d^3{\bf y} \left(\mathcal{U}(y) - \frac{\mathcal{U}(y)^2}{2\epsilon^2} + \frac{(\nabla \mathcal{U}(y))^2}{4\epsilon^4}\right).
\end{equation}
The first term is the standard LDA result and the other terms are the corrections in powers of $1/\epsilon^2$ on top of it. Note that this result is valid both for weak potentials and for potentials tuned to unitarity. We further expect this expression to hold for sufficiently smooth potentials, so that the gradient term can be treated perturbatively.

We can compute the number of particles in the polaronic cloud using the relation
$N = -\partial E/\partial \mu$, where the expression for energy should be first expressed in terms of $\mu$. 
This gives:
\be
\label{eq:NLDA}
N = -\frac{r_c}{8\pi a_b}\int d^3{\bf y} \,\left(\mathcal{U}(y) - \frac{(\nabla \mathcal{U}(y))^2}{4\epsilon^4}\right).
\ee

For the Pöschl-Teller potential given in Eq.~\rf{eq:PT} we have $\mathcal{U}(y) = -{2}/{\cosh^2(y)}$. This leads to 
\be
\label{eq:EPT}
E_\text{PT} = -\frac{\pi n_0 r_c}{m}\left(\frac{\pi^2}{3} + \frac{2(\pi^2-6)}{9\epsilon^2}  - \frac{4\pi^2}{45\epsilon^4} \right),
\ee
\be
\label{eq:NLDAPT}
N_\text{PT} = \frac{\pi^2}{12}\frac{r_c}{a_b}.
\ee

In the intermediate regime where $\epsilon \sim 1$, instead, we do not expect that there exists an analytic and universal (depending on a single parameter that describes the boson-impurity potential) solution to the problem. For example, we solved the GPe analytically for the case of a shallow square well potential for arbitrary values of $\epsilon$ and we showed that unless $\epsilon \ll 1$  the solution is sensitive to details of the potential. The details of the derivation are presented in the Appendix \ref{sec:weak}.

\begin{figure}[t]
\centering
\includegraphics[width=\columnwidth]{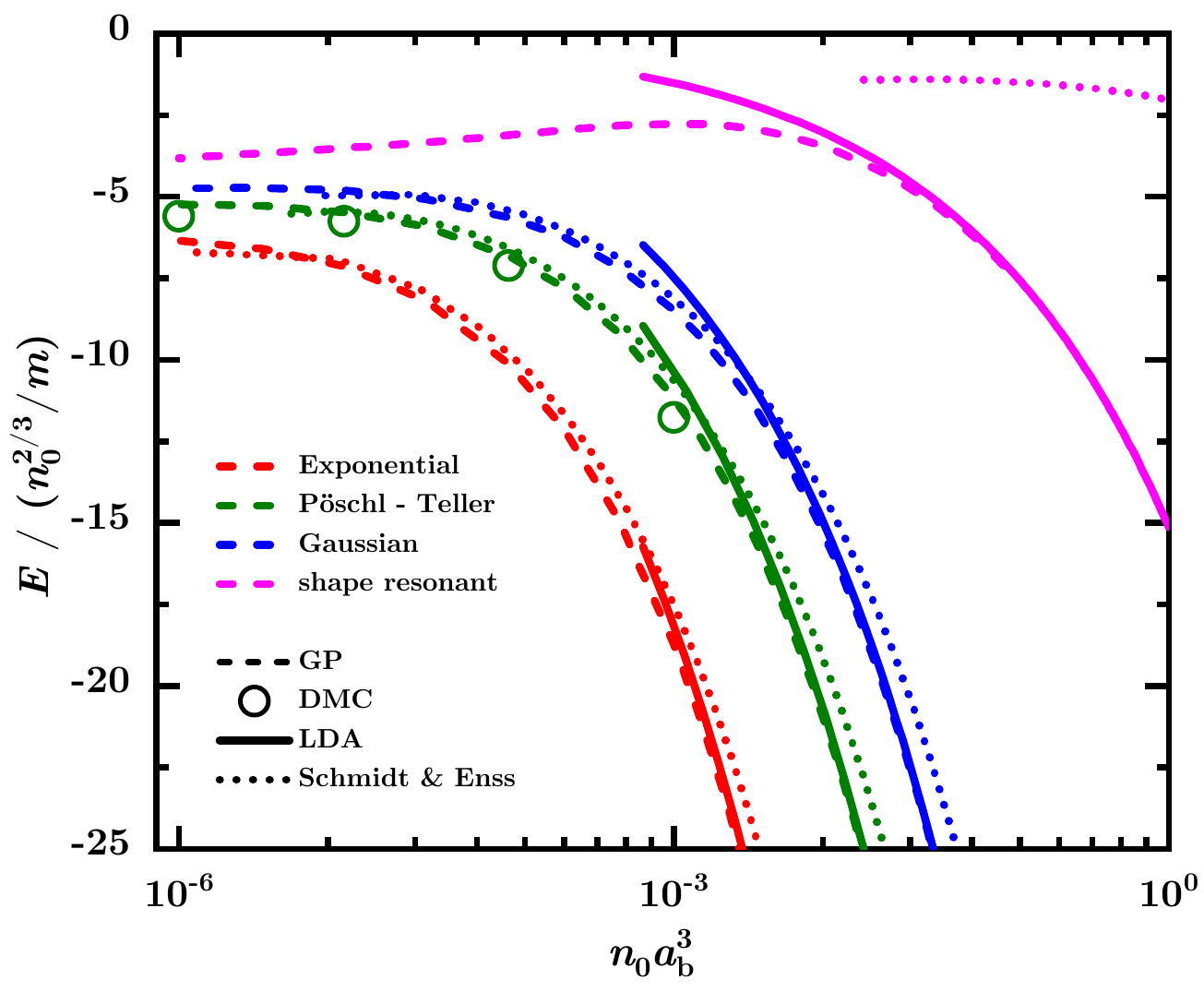}
\caption{
\label{fig:comparison_with_LDA_and_Schmidt_and_Enss}
Polaron energy as a function of the gas parameter for the ratio between the bath scattering length and the range of the potential equal to $a_b/\mathcal{R} = 0.1$.
Dashed lines represent the numerical solution of the local GPe~\eqref{eq:GPl}. 
Symbols show DMC data points. 
Solid lines at high density indicate the first term of the LDA prediction as summarized in Table~\eqref{table:LDA}.
Dotted lines depict the prediction from Eq.~\eqref{eq:S&E}, taken from Ref.~\cite{Schmidt2021}.
}
\end{figure}

As an illustration of the above analysis, in Table~\ref{table:LDA} we consider various unitary potentials $\mathcal{U}(y)$, and for each of those we compute the effective range $r_\text{eff}$ and the LDA expression for the polaron energy. 
The energies obtained via the GP equation, its LDA approximation, and QMC are shown in Fig.~\ref{fig:comparison_with_LDA_and_Schmidt_and_Enss}. The results of the local GPe are obtained using $r_b=0$, while in DMC calculations we use $r_b=r_c=\mathcal{R}$.

\begin{widetext}
\begin{center}
\begin{table}[h!]
\begingroup
\setlength{\tabcolsep}{12pt} 
\renewcommand{\arraystretch}{1.6} 
\begin{tabular}{ |c|c|c|c| }
\hline
Potential & $\mathcal{U}(y)$ at unitarity & $r_\text{eff}/r_c$ & LDA Energy [in units of $2\pi n_0 r_c/m$]  \\ 
\hline
Pöschl-Teller &$-\frac{2}{\cosh^2{y}}$ & 2 & $-\left(\frac{\pi^2}{6} + \frac{(\pi^2-6)}{9\epsilon^2}  - \frac{2\pi^2}{45\epsilon^4}\right)$ \\
Gaussian & $-2.65e^{-y^2}$ & 1.44 & $-\left(1.19+\frac{1.13}{2\epsilon^2}-\frac{3.39}{4\epsilon^4}\right)$\\
Exponential & $-1.45e^{-y}$ & 3.53 & $-\left(2.89+\frac{0.523}{2\epsilon^2}-\frac{0.523}{4\epsilon^4}\right)$ \\
Shape-resonant & $-3.73e^{-y^2}+0.707e^{-y}$ & 0.0336 & $-\left(0.241+\frac{1.47}{2\epsilon^2}-\frac{5.12}{4\epsilon^4}\right)$\\
Square well & $-\left(\frac{\pi}{2} \right)^2\Theta(1-y)$ & 1 & $-\frac{\pi^2}{12}$\\
 \hline
\end{tabular}
\endgroup
\caption{\label{table:LDA}
Effective ranges and LDA polaron energies for four different unitary potentials. 
The LDA may not be applied reliably to the square well, because the latter has a sharp discontinuity at $y=1$; as a consequence, for the square well we provide only the leading term in the energy.}
\end{table}
\end{center}
\end{widetext}


To conclude this Section, we compare our results with the ones reported by Schmidt and Enss in Ref.~\cite{Schmidt2021}. According to them, in the regime where $r_\text{eff} \geq 0.2\xi$ the energy of a unitary and infinitely-massive  polaron should be given by
\be
\label{eq:S&E}
E_\text{SE} = -\frac{n_0}{2m}\left[5.2(2)\xi + 9.0(1)r_\text{eff} \right].
\ee
When $r_\text{eff}\geq \xi$, the second term in the expression is dominant, giving $E\propto r_\text{eff}$.
At the same time, generally $r_\text{eff} \sim r_c$ and in the regime where $r_c \geq \xi$ we have shown that the LDA becomes applicable. 
This leads us to conclude that the energy should not scale with $r_\text{eff}$, but rather with $r_c$ [see Eq.~\eqref{eq:ELDA}]. Indeed, in Fig.~\ref{fig:comparison_with_LDA_and_Schmidt_and_Enss} we show that in the regime of large bath density (such that $\epsilon>1$) the numerical solution of the GPe agrees better with the LDA result than with Eq.~\rf{eq:S&E}. 
The discrepancy between the two approaches becomes particularly evident for so-called shape-resonant potentials, which are fine-tuned so that their effective range $r_\text{eff}$ vanishes.
In this case Eq.~\rf{eq:S&E} clearly fails, while the LDA result remains valid. 
For example, considering the shape-resonant potential given in the last line of Table \ref{table:LDA}, the energy obtained from the numerical solution of the GPe when $\epsilon=3$ agrees with the LDA result within $1\%$. 

\subsection{Accounting for a non-zero range in $V_{bb}$ via the generalized GPe}
\label{sec:ggpe}
\begin{figure}[b]
\centering
\includegraphics[width=\columnwidth]{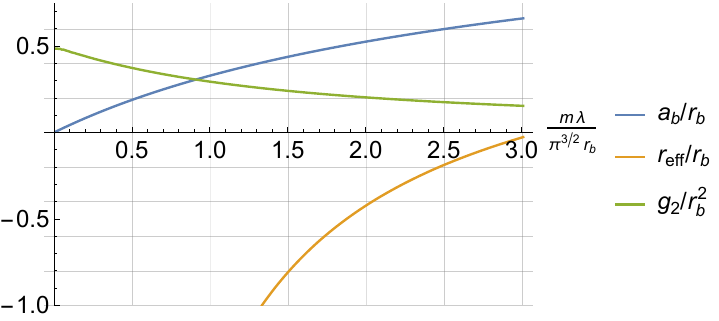}
\caption{
\label{fig:scattering_parameters_gaussian_potential}
Scattering length, effective range and coefficient $g_2$ for the repulsive gaussian potential defined in Eq.~\eqref{eq:vbb1}.
}
\end{figure}

To conclude our analytical considerations, we consider the effect of a non-zero range $r_b$ in $V_{bb}$. 
In our previous work~\cite{Yegovtsev2022} we demonstrated that as long as $r_b\lesssim r_c$, the solution of the non-local GPe with $\epsilon \ll 1$ still scales as $1/\epsilon^{2/3}$ close to the impurity.
As such, we expect that the analytical results obtained above using the local GPe will have a weak dependence on $r_b$ in the experimentally realistic scenario where $a_b\sim r_c \sim r_b$.
Another way to prove that the effects of a finite range between bath bosons are mild is to consider the generalized GPe introduced in Ref.~\cite{Collin2007}, which takes into account explicitly a non-zero effective range $r_{\rm eff}$ of the boson-boson potential $V_{bb}$:
\be
\label{eq:gGPl} \left[- \frac{\Delta }{2m} + U  -  \mu  + \lambda \bigg(| \psi({\bf x}) |^2 + g_2\Delta |\psi({\bf x})|^2\bigg) \right]  \psi({\bf x}) = 0,
\ee
with $g_2 = \left( \frac{a_b^2}{3}-\frac{a_b r_\text{eff}}{2}\right) $. For the case of a repulsive Gaussian potential, the coefficient $g_2$ turns out to be always positive and small, see Fig.~\ref{fig:scattering_parameters_gaussian_potential}. A similar behaviour is obtained also for a repulsive square well. As such, the effect of this correction can be safely estimated by first order perturbation theory.
Calling $\psi_0$ the solution of Eq.~\rf{eq:gGPl} with $g_2=0$, the shift in energy due to the new term is: 
\be
\label{eq:DeltaE} \Delta E = \frac{\lambda\,g_2}{2}\int d^3{\bf x}\,|\psi_0({\bf x})|^2\nabla^2 |\psi_0({\bf x})|^2 
\ee
A numerical evaluation of the latter expression proves that it is always negative, across the whole range of parameters explored, so this correction consistently lowers the total energy. Our numerical study also shows that the ratio $\Delta E/E$ is always small, typically of the order of a few percent, and compatible with the downshift we found by means of the non-local GPe.
In the following, we will show that this is precisely the case in the analytically-solvable limit $\delta\ll1$.

\subsubsection{Analytical results at low gas density}
Let us focus on the potentials considered in the main text: unitary impurity-bath Pöschl-Teller potential, for which  $v(r)=\tanh(r)$ and $R\approx r_c$, and a Gaussian repulsive potential between bath bosons, with ranges $r_b=r_c$. 
The PT potential does not vanish identically beyond some range. However, it goes to zero sufficiently fast beyond $r_c$, so that for example $U(r=3r_c)\approx-0.01/(mr_c^2)$. As such, to a very good approximation in the limit $\delta=R/\xi \ll 1$ the wave function in such potential is given by 
\be \label{eq:fullsolsPT}  \phi(r)  \approx \left\{  \begin{matrix}  \frac{\xi^{2/3} R^{1/3} }{r}  \, v\left(\frac r {r_c} \right), & r < 3r_c , \cr 1+\frac{\xi^{2/3} R^{1/3} }{r}   e^{- \frac{\sqrt{2} r}{\xi}},  & r>3r_c .   \end{matrix}
\right. 
\ee
To proceed, we write $\Delta E=\Delta E_{\rm in}+\Delta E_{\rm out}$, where each term corresponds to the contribution to the energy shift coming from the regions $r<3r_c$ and $r>3r_c$.
In the neighborhood of the impurity, $r<3r_c$, the wave-function \eqref{eq:fullsolsPT} decays rapidly (i.e., it has a large and negative curvature), and a numerical evaluation of the corresponding  contribution $\Delta E_{\rm in}$  gives
\be\label{eq:DeltaEIn}
\Delta E_{\rm in}
=-1.58\frac{n_0\xi}{m}\frac{g_2}{r_c^2} \frac{\xi}{r_c}\delta^{4/3}
,
\ee
which is always negative.
Its importance with respect to the unperturbed energy \eqref{eq:Eunit} is
\be
\frac{\Delta E_{\rm in}}{|E|}\approx -0.167 \frac{g_2}{r_c^2}\frac{R}{r_c}= -0.175 \frac{g_2}{r_c^2}.
\ee
Taking the maximum value of $g_2$ shown in Fig.~\ref{fig:scattering_parameters_gaussian_potential}, one sees that $\Delta E_{\rm in}$ can be at most $9\%$ of the local-GPE energy $E$.

\begin{figure}[t]
\centering
\includegraphics[width=\columnwidth]{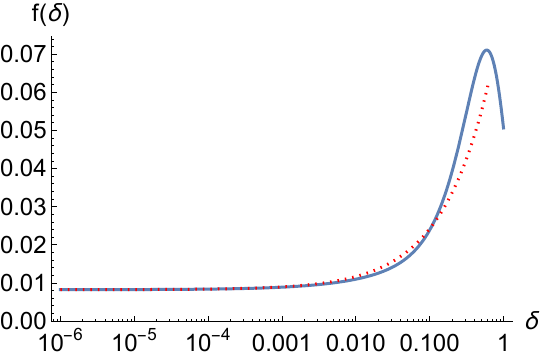}
\caption{
\label{fig:f_of_delta}
Function $f(\delta)$ defined in Eq.~\eqref{eq:DeltaEOut}. The dotted line shows its series expansion $f(\delta)=0.008+0.07\delta^{2/3}$.
}
\end{figure}

On the contrary, in the Yukawa tail away from the impurity the wavefunction has a small and positive curvature, and the explicit calculation of the integral in Eq.~\eqref{eq:DeltaE} for $r>3r_c$ shows that $\Delta E_{\rm out}$ is positive and very small:
\be\label{eq:DeltaEOut}
\frac{\Delta E_{\rm out}}{|E|}
= 
\frac{g_2}{\xi^2\delta^2}f(\delta)
\approx 
\frac{g_2}{r_c^2}f(\delta)
\ee
where $f(\delta)$ is the function shown in Fig.~\ref{fig:f_of_delta}. For small $\delta$ we have $f(\delta)\approx 0.008 + 0.07 \delta^{2/3}$, and as such we find $\Delta E_{\rm in}\approx -20\Delta E_{\rm out}$ for small $\delta$. 

In conclusion, we have shown explicitly that for small $\delta$ (i.e., for a dilute gas) the correction $\Delta E$ due to the non-zero range of $V_{bb}$ is small and negative, as anticipated, and it comes mainly from the region in the vicinity of the impurity, where the wavefunction varies rapidly.

\begin{figure}[t]
\centering
\includegraphics[width=\columnwidth]{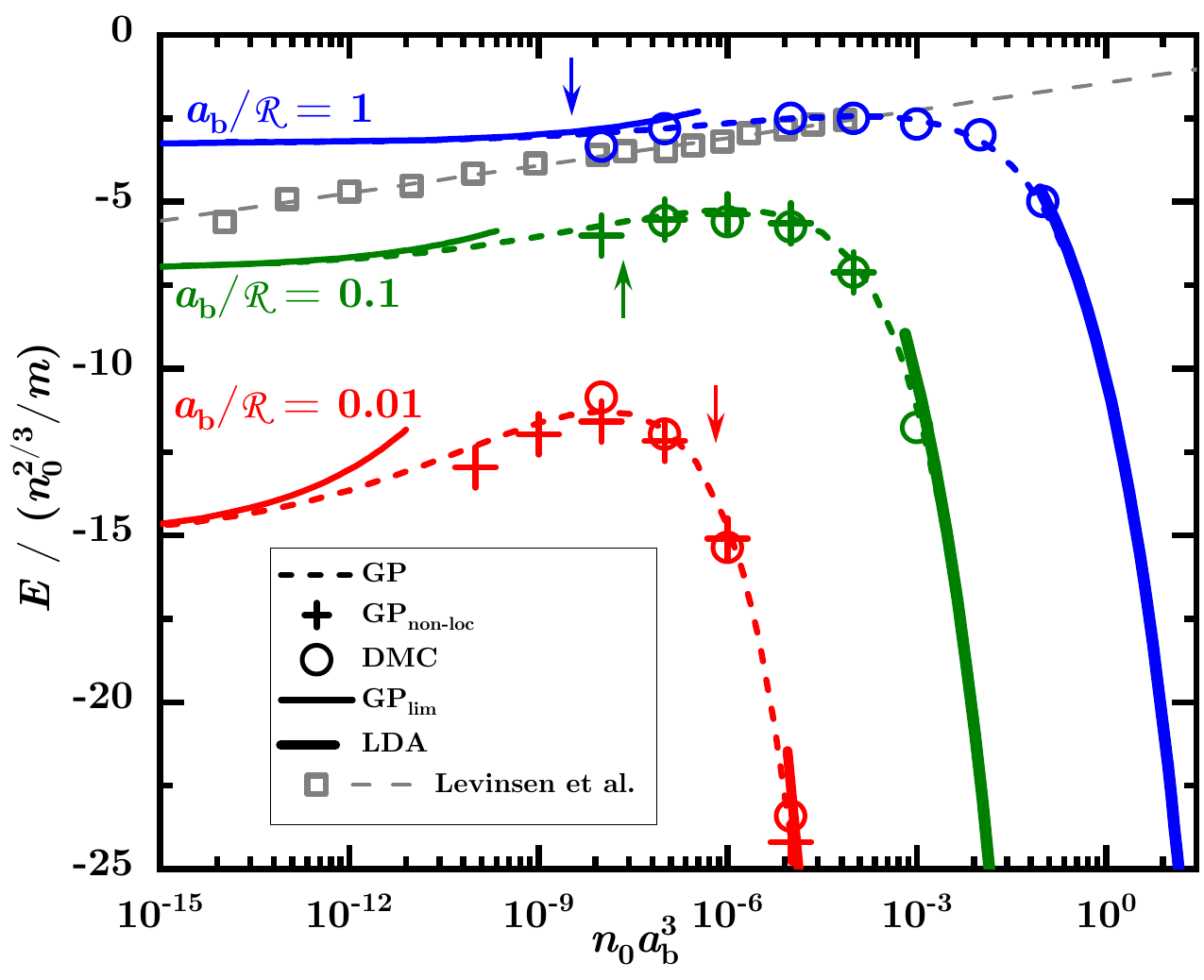}
\caption{\label{fig:energy_vs_gas_parameter}
Polaron energy as a function of the gas parameter for three values of the ratio between the bath scattering length $a_b$ and the range of the potentials $\mathcal{R}$: from top to bottom, $a_b/\mathcal{R} = 1, 0.1, 0.01$.
The dashed lines represent the numerical solution of the local GPe, Eq.~\eqref{eq:GPl}.
Plus symbols represent the data points obtained by solving the non-local GPe, Eq.~\eqref{eq:GPnl}. 
The circles show DMC results with the statistical error bars smaller than the symbol size. 
The arrows indicate the densities at which the polaron size is comparable with the box size for 100 particles, highlighting the onset of significant finite-size effects in DMC calculations for lower densities.
Thin solid lines at low density show the analytic prediction of Eq.~\eqref{eq:Eunit}, while the solid thick lines on the right side indicate the LDA prediction given in Eq.~\eqref{eq:EPT}.
The squares and grey dashed line show DMC data and their logarithmic fit from Ref.~\cite{Levinsen2021} (obtained considering $a_b=r_b$ but $r_c=0$).
}
\end{figure}
\section{Numerical results}
\label{sec:numerics}
Let us now switch to describing our numerical results. Please refer to the Appendix~\ref{sec:appxc} for the details of the numerical simulations.
We compute the energy of the unitary polaron for different values of $a_b$, comparing the results of three different methods: local GPe, non-local GPe, and DMC. 
In all cases, we consider the experimentally relevant situation of equal ranges for the potentials $V_{bb}$ and $U$, i.e., we set $r_b=r_c=\mathcal{R}$ in Eqs.~\rf{eq:vbb1} and \rf{eq:PT}. 
In the DMC calculations, the amplitude $\lambda$ of the Gaussian interaction potential~(\ref{eq:vbb1}) between bosons is chosen such that the scattering length obtained from the two-body Schrödinger equation is equal to $a_b$.
Instead, in the simulations of the local and non-local GPe, we fix $\lambda$ via the Born relation $\lambda=4\pi a_b/m$. 
We do so even when $a_b=\mathcal{R}$, where the Born approximation is clearly not satisfied, and the discrepancy with the exact scattering length is large (see Appendix \ref{sec:born} for details). 
As we will see below, this unorthodox choice still yields remarkably accurate polaron energies.

The energies obtained by the three methods are compared
in Fig.~\ref{fig:energy_vs_gas_parameter}.
The open circles denote DMC results, the dashed lines represent outcomes of the local-GPe~\rf{eq:GPl}, and plus-symbols depict results of the non-local GPe~\rf{eq:GPnl}.
The non-local GPe yields lower energies than the local-GPe, because the nonlocal potential is effectively softer than the contact one, and therefore more bosons can be accommodated within the polaron cloud surrounding the impurity. 
The numerical solution of the generalized Gross-Pitaevskii equation introduced in Ref.~\cite{Collin2007} yields a similarly small downshift of the energy.
For small gas parameters (left side of the graph), $\xi$ is large and our numerical energies from both GPe and DMC converge towards our analytical solution of the unitary polaron, Eq.~\rf{eq:Eunit} (thin solid lines). On the contrary, on the right side of the graph (where $\xi$ becomes smaller than $r_c$) our energies smoothly approach the LDA result~\rf{eq:EPT} (thick solid lines).

We find a remarkable agreement between DMC and the GPe for all considered values of $a_b/\mathcal{R}$.
The minor discrepancies observed between DMC and GP results at very low densities might be attributed to finite-size effects.
More specifically, finite-size effects in DMC start to become important for densities below the vertical arrows, which indicate where the local GPe predicts the number of particles in the polaron dressing cloud to be equal to the typical number of particles used in DMC calculations.
The number of particles in the dressing cloud of the polaron can be estimated using LDA, Eq.~\rf{eq:NLDAPT}.
As a result, the smallest gas parameter $n_0a_b^3$ that we were able to reliably investigate was achieved for the largest value of $a_b/\mathcal{R}$. 

Levinsen {\it et al.} studied a similar physical setup in Ref.~\cite{Levinsen2021}, choosing however a contact impurity-bath potential $U$ and a hard-sphere bath-bath potential $V_{bb}$ (i.e., they considered $r_c=0$ and $a_b=r_b$). 
Their DMC results and a logarithmic fit to them are shown in Fig.~\ref{fig:energy_vs_gas_parameter} with grey symbols and a dashed line, correspondingly. Despite the different choices of potentials, their DMC data are in remarkable agreement with our DMC and GPe results as long as the total number of particles in the polaronic cloud does not exceed 100, as is indicated by the blue arrow in  Fig.~\ref{fig:energy_vs_gas_parameter}. 
In the regime of very low gas parameter, however, their DMC energies are significantly lower than our GPe predictions. 
As discussed above, we have not been able to produce reliable DMC results at such extremely low densities due to pronounced finite-size effects.

\begin{figure}[t]
\centering
\includegraphics[width=\columnwidth]{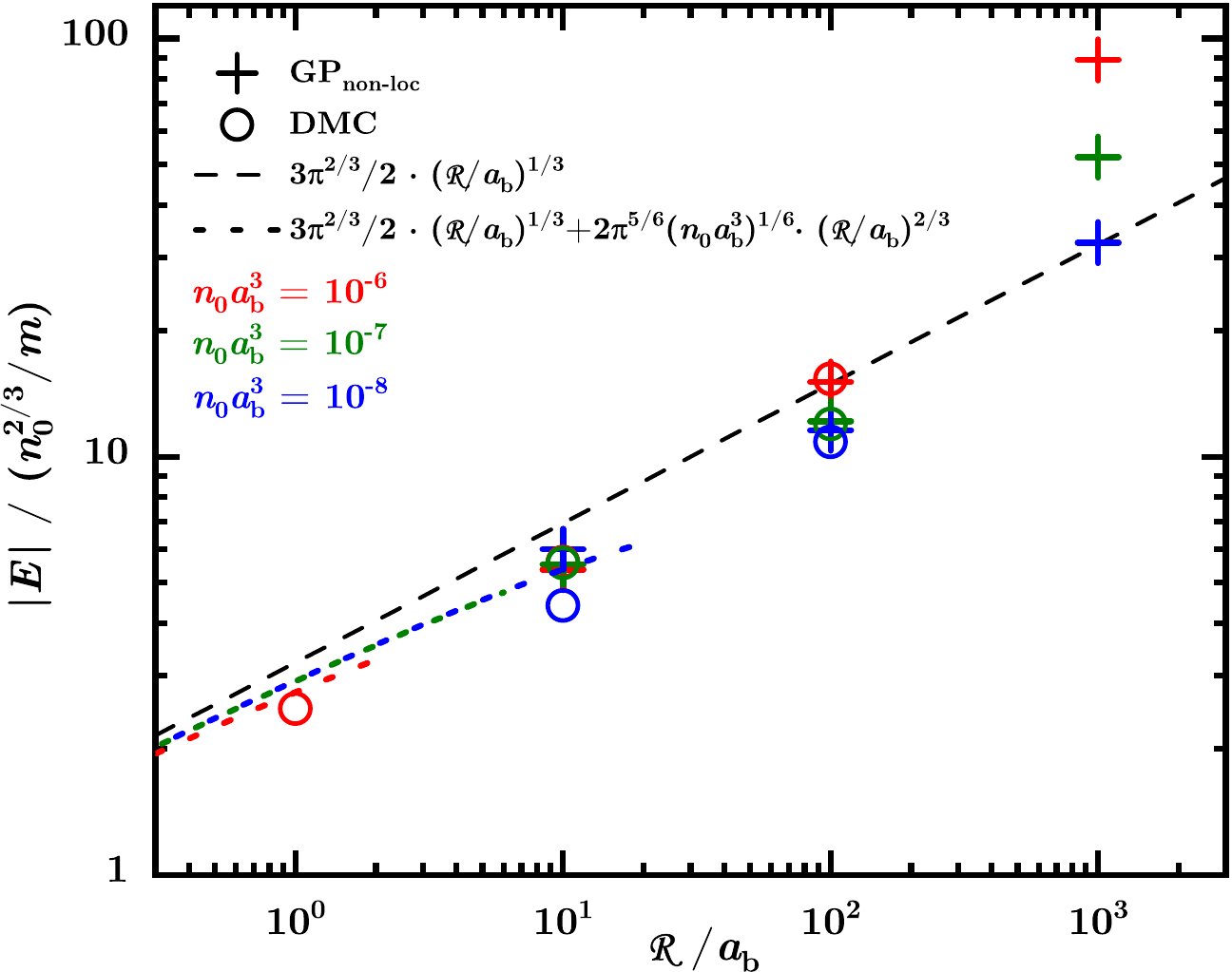}
\caption{\label{fig:energy_vs_range}
Polaron energy as a function of the range of the interactions of the bath particles for three characteristic values of the gas parameter $n_0a_b^3 = 10^{-6}, 10^{-7},10^{-8}$ represented by red, green and blue colors and ordered from top to bottom. 
Symbols:
pluses, numerical solution of the non-local GPE; 
circles, DMC results.
The dashed (dotted) line represents the leading term (two terms) of Eq.~(\ref{eq:Eunit}).
} 
\end{figure}

To investigate further the effects of non-zero range, in Fig.~\ref{fig:energy_vs_range} we examine the energy of the polaron as a function of $\mathcal{R}/a_b$ for three different values of the gas parameter $n_0a_b^3$. To avoid finite-size effects, DMC points are shown only for the values of the gas parameter for which the number of particles in the polaron dressing cloud is smaller than the total number of particles in our simulations, i.e., 100. 
By changing the value of $\mathcal{R}/a_b$ on the horizontal axis of Fig.~\ref{fig:energy_vs_range}, we move between the different curves depicted in Fig.~\ref{fig:energy_vs_gas_parameter}. 
Figure~\ref{fig:energy_vs_range} illustrates that in a dilute bath for sufficiently small value of the potential range, the polaron energies follow the asymptotic law $E\propto n_0^{2/3}(\mathcal{R}/a_b)^{1/3}$ (dashed line) with the subleading term given by Eq.~\rf{eq:Eunit}. 
Note that the latter equation was derived for $r_b=0$, but our numerical results shown here are for ${\cal R} = r_b=r_c$. This proves that a non-zero range of $V_{bb}$ yields negligible changes to the energy (as we had argued above, at the end of our analytical considerations).
The condition for the applicability of this law, $\mathcal{R}\ll \xi$, can be expressed equivalently 
as $\sqrt{8\pi (n_0a_b^3) (\mathcal{R}/a_b)^2} \ll 1$. Focusing on the rightmost points in Fig.~\ref{fig:energy_vs_range} (the ones with $\mathcal{R}/a_b = 10^3$) the condition $\mathcal{R}\ll \xi$ is clearly violated for $n_0a_b^3 = 10^{-6}$ (red cross, where $\mathcal{R}\approx 5\xi$), and a clear departure from the trend $E\propto a_b^{-1/3}$ is observed. However, for a more dilute gas with $n_0a_b^3 = 10^{-8}$ (blue cross) one has $\mathcal{R}\approx \xi/2$, and the scaling $E\propto a_b^{-1/3}$ is approached.

\section{Conclusions}
\label{sec:conslusions}
In conclusion, we analyzed the Bose-polaron problem at unitarity by means of the DMC and the GP (local, and non-local) methods. Our study revealed a remarkable agreement between these approaches in experimentally-relevant conditions. We also showed that the GP results remain accurate even when the microscopic boson-boson potential does not satisfy the Born approximation and all ranges of the problem are comparable to each other. 
Moreover, we showed that including a non-zero boson-boson range yields negligible corrections to the polaron energy, which implies that for many practical purposes the analysis based on a local version of the GPe is sufficient for making qualitative and quantitative predictions for polarons all the way between weak-coupling and unitarity. 
We have also shown that there are two regimes, 
obtained when either $r_c \ll \xi$ or when $r_c \gg \xi$, 
where the problem admits an analytical solution (which is universal in the first case, but potential-dependent in the second).
In the intermediate regime where $r_c \sim \xi$ the properties of Bose polarons are not expected to be universal, even in the weak coupling limit. 
In particular, this may explain the discrepancy between the recent Monte Carlo study~\cite{astrakharchik2023many} and the perturbative analysis of the $r^{-4}$ polarization potential~\cite{ding2022mediated}, where the characteristic range of the polarization potential $R^*$ was of the same order as the healing length $\xi$. 
Indeed, the analysis based on many-body perturbation theory seems to be reliable only in the regime where $r_c \ll \xi$ and $|a|\ll (R/a_b)^{1/3} n_0^{-1/3}$ \cite{Massignan2021}, where one can expand the fields around the ground state of the weakly interacting BEC without the impurity. The overall remarkable agreement between GPe results and exact DMC calculations demonstrated here proves that the GPe is a reliable and ideal framework for predicting Bose polaron energies. An open question is whether such good agreement extends to other properties. Some of us showed earlier that the GPe can also be used to obtain analytical expressions for the polaron's effective mass and the induced interactions between two unitary Bose polarons~\cite{PhysRevA.108.L051301}. We plan to verify those results in a forthcoming DMC study. Furthermore, the GPe shall provide important guidance towards characterizing the orthogonality catastrophe in a Bose gas, which has been predicted long ago but not yet observed.


\vspace{5mm}
\begin{acknowledgments}
We acknowledge insightful discussions with G.~Bruun, J. Levinsen and M. Parish.
This work was supported by the Simons Collaboration on Ultra-Quantum Matter, which is a grant from the Simons Foundation (651440, VG, NY), and by the National Science Foundation under Grant No. NSF PHY-1748958.
G.E.A.~and P.M.~acknowledge support by the Spanish Ministerio de Ciencia e Innovación (MCIN/AEI/10.13039/501100011033, grant PID2020-113565GB-C21), and by the Generalitat de Catalunya (grant 2021 SGR 01411). 
P.M.~further acknowledges support by the {\it ICREA Academia} program, the Institut Henri Poincaré (UAR 839 CNRS-Sorbonne Université) and the LabEx CARMIN (ANR-10-LABX-59-01).
\end{acknowledgments}
\appendix
\section{Applicability of the Born approximation}
\label{sec:born}
The Gross-Pitaevskii equation (GPe) is generally derived under the approximation that the potential $V_{bb}({\bf x})$ satisfies the criteria of the applicability of the Born approximation.
To see this we can look for a spatially uniform solution of the 
GPe without the polaron potential. Substituting $\psi=\psi_0$ into the non-local GPe~\rf{eq:GPnl} with $U=0$ we find 
\be
\mu =  \left| \psi_0\right|^2 \int d^3{\bf x} \, V_{bb}({\bf x}).
\ee
Comparing with the standard equation of state of a weakly interacting Bose gas, $\mu = 4 \pi  a_b \left| \psi_0 \right|^2/m$ immediately gives
\be \label{eq:abBorn} a_b =\frac{m}{4 \pi} \int d^3{\bf x} \, V_{bb}({\bf x}).
\ee
This is nothing but the expression given by the Born approximation for the scattering length $a_b$ in the potential $V_{bb}$. It is well known that Eq.~\rf{eq:abBorn} works for those potentials whose range $r_b \gg a_b$, therefore Eq.~\rf{eq:GPnl} strictly-speaking holds only under this condition. 

Nonetheless, the GPe is used routinely to describe experiments with ultracold dilute Bose gases, where typically $r_b \sim a_b$.
One could therefore reasonably ask if the Gross-Pitaevskii equation is entirely incompatible with potentials whose scattering length is not given by the Born approximation expression \rf{eq:abBorn}. 
It is not difficult to see that in principle it is  possible to use GPe even when the potential in it does not obey Eq.~\rf{eq:abBorn}, that is even if $a_b \gtrsim r_b$. 
Formally, in this case the potential $V_{bb}({\bf x})$ in Eq.~\rf{eq:GPnl} must be replaced by the so-called vertex function, computed up to all orders of perturbation theory. The usual textbook approach to the Gross-Pitaevskii equation, valid when GPe does not include the boson-impurity potential $U$, is to argue that the solutions to the GPe are not sensitive to the dependence of the potential $V_{bb}$ on the coordinates and replace the potential by the delta-function potential given by \rf{eq:GPl}, whose strength $\lambda$ is adjusted to produce the desired scattering length $a_b$ according to Eq.~\rf{eq:abBorn}. However, since our GPe does include the potential $U$ and its solutions depend on the range of that potential, it is not clear {\sl a priori} whether the solutions to it depend on the range of $V_{bb}({\bf x})$ as well. In fact, in Ref.~\cite{Yegovtsev2022} we examined this question and found that the solutions to Eq.~\rf{eq:GPnl} depend on the range of the potential $V_{bb}$ only mildly. 

 In this work we take a practical approach to this problem. If $a_b \ll r_b$ we use the equation \rf{eq:GPnl} as is. If $a_b \gtrsim r_b$, we use the simplified equation \rf{eq:GPl} with the properly adjusted $\lambda$, counting on this equation still providing a reasonable approximation to the solution that we seek. 
 Quite remarkably, as we show in the main text, the polaron energies obtained by the GPe with this choice of $\lambda$ agree closely with the ones obtained by exact DMC calculations.

For the  interaction potential $V_{bb}$ given by the Gaussian potential in Eq.~\rf{eq:vbb1}, the table below compares the exact values of $a_b$ obtained from the numerical solution of the Schrödinger equation for various values of $\lambda$ with $a_b$ computed using the first Born approximation Eq.~\rf{eq:abBorn}. As should be expected, the validity of the first Born approximation gets worse with increasing $a_b/r_b$.

\begin{table}[!h]
\begin{center}
\begin{tabular}{ |c|c|c| }
\hline
$a_b$ [Born, Eq.~\rf{eq:abBorn}] & $a_b$ [exact] & Relative error $(\%)$  \\ 
\hline
 0.01 & 0.00992 & 0.792 \\
 0.1 & 0.0926 & 7.95 \\
 3.63 & 1 & 263 \\
 \hline
\end{tabular}
\caption{Value of the boson-boson scattering length $a_b$ [in units of the boson-boson potential range $r_b$], obtained varying the strength $\lambda$ of the potential $V_{bb}$.}
\end{center}
\end{table}
\section{Bose polaron at weak coupling}
\label{sec:weak}
Here we present a toy model which shows that when $r_c \sim \xi$ the solution of the GPe will generally depend on various details of the potential. As such, a universal description of the Bose polaron is not expected to exist in this regime even in the weak coupling limit. A shallow square well potential may be written as $U(r) = \frac{1}{2mr_c^2}\left(\frac{\alpha\pi}{2}\right)^2\Theta(1-y)$ with $\alpha \ll1$ ($\alpha=1$ corresponds instead to the unitary point). The corresponding dimensionless GPe reads:
\begin{equation}
-\nabla^2\phi + \epsilon^2(\phi^2-1)\phi = \left(\frac{\alpha\pi}{2}\right)^2\Theta(1-y)\phi.
\end{equation}
Since $\alpha$ is small, we may seek a solution in the form $1+\delta$:
\begin{equation}
-\nabla^2\delta + \epsilon^2(2\delta+3\delta^2+\delta^3) =  \left(\frac{\alpha\pi}{2}\right)^2\Theta(1-y)(1+\delta).    
\end{equation}
Neglecting the nonlinear terms and writing $\delta = \frac{u}{y}$, we get:
\begin{equation}
-u'' + \left(2\epsilon^2 -  \left(\frac{\alpha\pi}{2}\right)^2\Theta(1-y) \right)u =  \left(\frac{\alpha\pi}{2}\right)^2y\Theta(1-y).  
\end{equation}
In the region $y>1$, we get $u=Be^{-\sqrt{2}\epsilon y}$. When $y<1$ we have three scenarios: 1) $2\epsilon^2> \left(\frac{\alpha\pi}{2}\right)^2$, 2) $2\epsilon^2 =  \left(\frac{\alpha\pi}{2}\right)^2$ and 3) $2\epsilon^2 <  \left(\frac{\alpha\pi}{2}\right)^2$. Below we consider only the first and the third scenarios, which are of the most physical relevance.

Let us focus on the first scenario. Defining $ \left(\frac{\alpha\pi}{2}\right)^2 = \omega_0^2$ and $2\epsilon^2 -  \left(\frac{\alpha\pi}{2}\right)^2 = \omega^2$, solution reads: $u = A\sinh{(\omega y)} + \frac{\omega_0^2}{\omega^2}y$. Matching the amplitude and the derivative at $y=1$, we get:
\begin{equation}
\begin{split}
& 1+ A\sinh{(\omega)} + \frac{\omega_0^2}{\omega^2} = 1 + Be^{-\sqrt{2}\epsilon},\\
& 1+ A\omega\cosh{(\omega)} + \frac{\omega_0^2}{\omega^2} = 1-\sqrt{2}B\epsilon e^{-\sqrt{2}\epsilon}.
\end{split}    
\end{equation}
Solving the above system one gets:
\begin{equation}
\begin{split}
& B = \frac{\omega_0^2}{\omega^2}\left(1-\frac{\tanh{(\omega)}}{\omega} \right)\frac{e^{\sqrt{2}\epsilon}}{\left(1+\frac{\sqrt{2}\epsilon}{\omega}\tanh{(\omega)} \right)},    \\
& A = \left(Be^{-\sqrt{2}\epsilon}-\frac{
\omega_0^2}{\omega^2}\right)\frac{1}{\sinh{(\omega)}}. 
\end{split}    
\end{equation}
This result depends on the details of the potential, however if we consider the limit $\xi \gg r_c$, or equivalently $\epsilon \ll 1$, we can simplify it further.  For a fixed $\alpha$ taking $\epsilon \rightarrow 0$ limit pushes the solution into the third regime $2\epsilon^2 < \left(\frac{\alpha\pi}{2}\right)^2$, where instead of a real $\omega$, we have an imaginary one: $\omega \rightarrow i\omega$, $\tanh{(\omega)}\rightarrow i\tan{(\omega)}$, and $\omega = |2\epsilon^2-\left(\frac{\alpha\pi}{2}\right)^2|^{1/2}$. Taking the limit $\epsilon \to 0$ gives:

\begin{equation}
\begin{split}
&\lim_{\epsilon \to 0} \left[\frac{-\omega_0^2}{\omega^2}\left(1 - \frac{\tan{(\omega)}}{\omega} \right) \frac{e^{\sqrt{2}\epsilon}}{\left(1+\frac{\sqrt{2}\epsilon}{\omega}\tan{(\omega)} \right)} \right]   =\\
&-\left(1-\frac{\tan{(\omega_0)}}{\omega_0} \right) = -a/r_c.
\end{split}
\end{equation}
In the last step we used the analytical formula for the scattering length of the square well potential. The solution inside the impurity potential becomes $\sinh{(\omega y)}\to i\sin{(\omega y)}$. When $\epsilon \to 0$, the whole solution inside the impurity potential becomes:
\begin{equation}
\begin{split}
&\lim_{\epsilon \to 0}\phi = \lim_{\epsilon \to 0}\left(1 + A \frac{\sin{(\omega y)}}{y \sin{(\omega)}} - \frac{\omega_0^2}{\omega^2}\right) = \\
&\left(1-\frac{a}{r_c} \right)\frac{\sin{(\omega_0 y)}}{y \sin{(\omega_0)}} =\frac{r_c}{r} \left(1-\frac{a}{r_c} \right) v\left(\frac{r}{r_c}\right).
\end{split}
\end{equation}
Here $v(r/r_c)$ is the solution to the zero energy Schrödinger equation that satisfies $v(0)=0$ and $v(1)=1$. This reproduces the result already quoted in Eq.~\rf{eq:fullsolw}. This analysis shows that the solution to the GPe has universal features (depends only on a single parameter such as scattering length $a$ or range $R$) only in the $\xi \gg r_c$ limit, but when those two length scales are of the same order it will depend on the details of the potential in some nontrivial way. If $r_c\gg \xi$, then one can solve the problem using the LDA discussed above.


\section{Details of the numerical simulations}
\label{sec:appxc}
\subsection{Non-local GPe} 
We find the ground state solution of the non-local GPe in Eq.~\rf{eq:GPnl} by solving it in imaginary time:
$$ -\partial_\tau\psi = \left( - \frac{\Delta }{2m} + U  -  \mu  + \int d^3 {\bf y} \ V_{bb}({\bf x}-{\bf y}) \left| \psi({\bf y}) \right|^2 \right)  \psi({\bf x}) 
$$
For numerical convenience, we introduce the variable $\phi(y) = \psi(y)/\sqrt{n_0}$ where $y=r/r_c$. We study the problem on a finite interval $r\in [0, \Lambda]$. At very large distances from the impurity we impose the Yukawa boundary condition: $\phi(y) =  1 + \frac{C}{r}e^{-\sqrt{2}\frac{r_c}{\xi}y}$. We use the method of lines by discretizing the space variable to obtain a system of coupled nonlinear differential equations in imaginary time. This method is very efficient for the case of local $V_{bb}$, where after finding the coefficient of the tail $C$, one can also add the contribution from the interval $y\in [\Lambda, \infty]$. For the nonlocal $V_{bb}$  this method introduces some artificial boundary effects near $r = \Lambda$. If $\Lambda$ is large enough, those effects do not change the behavior of the solution in the bulk. In order to compute the energy, we choose the size of the interval to be large enough, so that the energy can be computed by using the solution on a smaller interval $y \in [0, \Lambda_0]$, $\Lambda_0<\Lambda$, such that the changes in both $\Lambda$ and $\Lambda_0$ produce little change in the energy of the polaron.

\subsubsection{Boson-boson interaction potential}
For the choice of potential in Eq.~\rf{eq:vbb1}, one can perform the angular integration in the expression inside the integral analytically and obtain:
\be \label{eq:symme}
\begin{split}
&\int d^3{\bf y} \, V_{bb}({\bf x}-{\bf y}) \left| \psi(\bf {y}) \right|^2 = \\
&\frac{\lambda }{\sqrt{\pi} r_b x}\int_0^\infty y\, dy \, e^{-\frac{(x-y)^2}{r_b^2}}\left(1-e^{-\frac{4xy}{r_b^2}} \right) |\psi(y)|^2. 
\end{split}
\ee 
When we discretize space, for every point $x$ we sum over all neighboring sites that are within a distance corresponding to 5 widths of the corresponding Gaussian. 

\subsection{Details of DMC simulations. }

The microscopic Hamiltonian of our model is given by
\be 
H =-\sum\limits_{i=1}^{N_b} \frac{\Delta_i}{2m} 
+ \sum\limits_{i<j}^{N_b} V_{bb}({\bf x}_i-{\bf x}_j) + \sum\limits_i^{N_b} U({\bf x}_i-{\bf X}),
\ee
where ${\bf x}_i$ represents the coordinates of the bosons and ${\bf X}$ denotes the position of the impurity. 
The DMC method is based on solving the many-body Schrödinger equation in imaginary time and allows one to find the ground-state energy exactly.
Simulations are performed for a system consisting of $N_b$ bosons (typically we use $N_b=100$) and a single impurity within a box of dimensions $L\cdot L\cdot L$ with periodic boundary conditions, which help to reduce the finite-size effects. 
The size of the simulation box is determined from the number of bosons $N_b$ and the average density $n_0$, such that $n_0=N_b/L^3$.

In order to reduce the statistical noise, an importance sampling technique is employed. 
The guiding wave function is chosen in the Jastrow pair-product form
\begin{equation}
\label{eq:Jastrow}
\Psi_T({\bf x}_1,\cdots,{\bf x}_N; {\bf X})
=
\prod\limits_{i<j}^{N_b}f_{bb}(|{\bf x}_i - {\bf x}_j|)
\prod\limits_{i=1}^{N_b}f_{bi}(|{\bf x}_i - {\bf X}|)\;.
\end{equation}
The boson-boson $f_{bb}(r)$ Jastrow terms are determined by solving numerically the two-body scattering equation for two bosons of mass $m$ interacting with the repulsive Gaussian $V_{bb}$ interaction potential in the range $0<r<L/2$.
The scattering energy is chosen such that the Jastrow term has zero derivatives at the borders, 
$f_{bb}'(0) = f_{bb}'(R_{L/2}) = 0$.
The boson-impurity $f_{bi}(r)$ Jastrow terms are obtained by solving the scattering problem for bosons of mass $m$ and a pinned impurity, interacting via an attractive Pöschl-Teller $U(r)$ potential in the range $0<r<R_{\rm match}$ with $f_{bi}'(0) = f_{bi}'(R_{\rm match}) = 0$.
The matching distance $R_{\rm match}$ is treated as a variational parameter that indirectly controls the value of the boson-impurity Jastrow term at zero, $f_{bi}(0)$. 
Additionally, it influences the boson-impurity pair distribution function and, consequently, the number of bosons in the polaron in the variational problem.
The specific values of the matching distance $R_{\rm match}$ are obtained by minimizing the total energy.

To determine the properties of a polaron immersed in a homogeneous system, we perform simulations for $N_b$ bosons contained in a box with periodic boundary conditions. This procedure induces finite-size effects on the polaron energy. A physically relevant parameter in this context is the ratio of the box size $L$, determined by the density $n_0 = N_b/L^3$, to the healing length $\xi$, which corresponds to the typical size of the polaron. This ratio can be expressed as 
$\frac{L}{\xi} = \sqrt[3]{8 N_b} (n_0 a_b^3)^{1/6}$
in terms of the number of bosons $N_b$ and the gas parameter $n_0 a_b^3$.

\begin{figure}[t]
\centering
\includegraphics[width=\columnwidth]{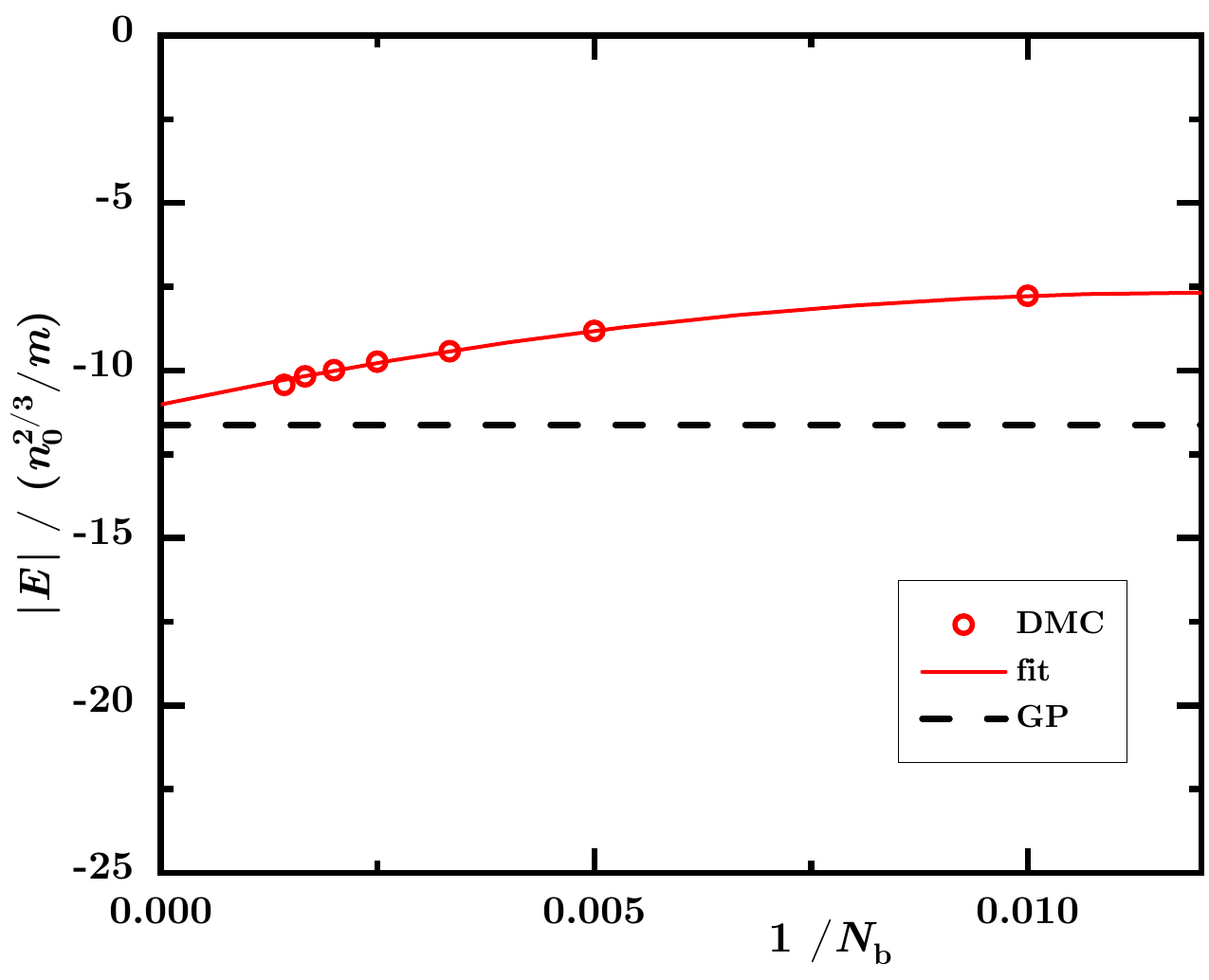}
\caption{
\label{fig:finite_size}
Polaron energy as a function of the inverse number of particles for the ratio between the bath scattering length and the range of the potential equal to $a_b/\mathcal{R} = 0.01$ and gas parameter $n_0a_b^3 = 10^{-9}$.
Symbols, DMC data points;
solid line, fit of type $f(x) = c_0 + c_1 x + c_2 x^2$;
dashed line, the numerical solution of the local GPe~\eqref{eq:GPl}. 
The vertical range is kept the same as in Fig.~\ref{fig:energy_vs_gas_parameter}.
}
\end{figure}

Figure~\ref{fig:finite_size} illustrates the dependence of the polaron energy on the inverse number of bosons, so that the thermodynamic value of the energy is obtained in the $1/N_b\to 0$ limit. 
The example shown in the figure corresponds to a very small gas parameter value, $n_0 a_b^3 = 10^{-9}$, for which finite-size effects are significant since the condition $L > \xi$ requires $N > 250$ particles in that case. 

To reduce the gas parameter further to $n_0 a_b^3 = 10^{-15}$ (the minimum value for which we report GPE results in Fig.~\ref{fig:energy_vs_gas_parameter}), the gas parameter must be diminished further by a factor of $10^6$. Maintaining the same value of $L/\xi$ ratio would require increasing the number of bosons by a factor of 1000. 
The Jastrow wave function~(\ref{eq:Jastrow}) contains $N_b^2$ terms, and their evaluation would increase the calculation time by a factor of $10^6$. 
This renders such calculations unfeasible with the current implementation of the DMC method. As a result, we limit the DMC results to parameters ($L\gtrsim\xi$ for $N_b=100$) for which accurate DMC calculations can be performed within a reasonable time.

\bibliography{UnitaryPolaron}

\onecolumngrid
\end{document}